\documentclass[aps,prc,reprint,superscriptaddress,nofootinbib]{revtex4-1}
\usepackage{mathrsfs}
\usepackage{amssymb}
\usepackage{graphicx}
\usepackage{dcolumn}
\usepackage{bm}
\usepackage{graphicx}
\usepackage{longtable}
\usepackage{amsmath}
\usepackage{color}
\usepackage{diagbox}
\usepackage{multirow}
\usepackage{braket}
\usepackage{booktabs}
\usepackage[colorlinks=true,linkcolor=blue,filecolor=blue, urlcolor=blue,citecolor=blue]{hyperref}
\usepackage{bookmark} 
\usepackage[colorlinks=true,linkcolor=blue,filecolor=blue, urlcolor=blue,citecolor=blue]{hyperref}
\usepackage{threeparttable}
\usepackage{amsmath}
\usepackage[normalem]{ulem} 
\usepackage{bbold}
\usepackage{appendix}
\usepackage{tensor}

\newfont{\largemi}{cmmi10}
\newfont{\smallmi}{cmmi6}

\draft

\def\eqref#1{Eq.~(\ref{#1})}

\fontsize{20}{20}

\begin{document}
	
\title{Pfaffian and minor formulas in number-conserved pairing theory}

\author{G. J. Fu}
\affiliation{School of Physics Science and Engineering, Tongji University, Shanghai 200092, China}

\date{\today}

\begin{abstract}

In this work, I present closed-form formulas for the norm and many-body density matrices between wave functions with exact particle numbers in pairing theory. The expressions, written as sums of Pfaffians and minors, apply to both even and odd particle-number systems and to pair condensates as well as broken-pair configurations. This formalism facilitates calculations using the generator coordinate method and symmetry restoration techniques, such as angular momentum projection. I apply these formulas to $^{70}$Ge and $^{69}$Ge using the angular-momentum-projected number-conserved BCS theory. The calculations describe low-lying collective bands and provide an interpretation of their intrinsic structures.

\end{abstract}
	
\vspace{0.4in}
	
\maketitle

\section{\texorpdfstring{Introduction}{Introduction}}

Pairing theories, such as the Bardeen-Cooper-Schrieffer (BCS) theory and the Hartree-Fock-Bogoliubov (HFB) approach, have been widely used in theoretical studies of quantum many-body systems.
While BCS and HFB approaches break U(1) symmetry, this violation has a minimal impact in large systems treated within the grand canonical ensemble.
However, for finite-size systems, such as atomic nuclei, ultra-cold atoms, and ultrasmall superconducting grains, 
particle-number conservation is crucial for accurately capturing intrinsic properties and quantum phase transitions.

To restore particle-number conservation, one common approach is particle number projection, which involves numerical integration over the gauge angle \cite{peterring,Dietrich1964}.
Alternatively, methods that directly conserve particle number offer a more straightforward description, for example, shape evolution and quantum phase transitions in rare-earth nuclei have been successfully described using number-conserved pair condensates in a unified framework~\cite{PNBCS2025Letter,PNBCS2025RareEarth}.
An $N$-pair condensate is defined as:
\begin{eqnarray}
	\ket{\rm{c}} \equiv \left( \hat{P}^{^{\dagger}} \right)^N  \ket{} , \label{pc}
\end{eqnarray}
where $N$ is the number of pairs, $\ket{}$ is the vacuum state, and $\hat{P}^{^{\dagger}}$ is the Cooper pair creation operator defined in a canonical basis of dimension $\omega = 2\Omega$ as 
\begin{eqnarray}
\hat{P}^{^{\dagger}} \equiv \sum_{\alpha =1}^{ \Omega} v_{\alpha} \hat{c}^{\dagger}_{\alpha} \hat{c}^{\dagger}_{\tilde{\alpha}}, \label{cooper}
\end{eqnarray}
where $v_{\alpha}$ is the pair coefficient.
The energy of the pair condensate can be computed based on recursive formulas derived from the generalized Wick theorem and the commutation relations between Cooper pairs \cite{talmibook,talmi1,npa1,npa2,PhysRevC.80.044335,PhysRevC.96.034313,PhysRevC.99.014302,npam1,npam2,PhysRevC.105.034317,npam3}. 
Nevertheless, computations in number-conserved pairing theory are generally complex.

Angular-momentum projection is required when the BCS or HFB wave function does not have good angular momentum.
This is done by integration over the Euler angles or by linear algebra projection~\cite{PHF1,PHF2}.
For instance, for angular momentum projection onto a pair condensate, the Hamiltonian and norm matrices can be written as
\begin{equation} 
	\begin{aligned}
		{\mathcal H}^J_{K'K} =& \frac{2J+1}{8\pi^2}\int {\rm d}\Omega ~ D^{J*}_{K'K}(\Omega) \langle {\rm c} | \hat{H} \hat{R}(\Omega) | {\rm c} \rangle		, \\
		{\mathcal N}^J_{K'K} =&\frac{2J+1}{8\pi^2}\int {\rm d}\Omega ~ D^{J*}_{K'K}(\Omega) \langle {\rm c} | \hat{R}(\Omega) | {\rm c} \rangle		,   \label{normH}
	\end{aligned}
\end{equation} 
where $D^J_{K'K}(\Omega)$ is a Wigner-$D$ matrix element and $\hat{R}(\Omega)$ is the rotation operator.

This work presents formulas for evaluating norms and matrix elements of many-body operators between wave functions with exact particle numbers, valid for both even and odd particle numbers. The formulas include configurations with broken pairs and matrix elements between different pair condensates in different canonical bases. They can therefore be used for configuration mixing in the generator coordinate method.

\section{\texorpdfstring{Formalism}{Formalism}}

Starting from Eq. (\ref{pc}), I extend the formalism to a more general broken-pair state with $n$ particles in a canonical basis:
\begin{eqnarray}
	\ket{\alpha} \equiv \left( \hat{P}^{^{\dagger}} \right)^N \hat{c}^{\dagger}_{\alpha_1} \cdots \hat{c}^{\dagger}_{\alpha_{\nu}} \ket{}, \label{bp}
\end{eqnarray}
where $n=2N+\nu$,
and $\nu$ is the number of unpaired particles, typically referred to as the seniority number in nuclear physics \cite{racah1942,racah1943}.
Eq. (\ref{bp}) applies to both even and odd particle numbers.
The canonical basis can be expressed as a unitary transformation of an orthonormal reference basis:
\begin{eqnarray}
	\hat{c}^{\dagger}_{\alpha}  = \sum_{i}  U_{\alpha i} \hat{a}^{\dagger}_{i}  , \label{unitaryU}
\end{eqnarray}
where $U$ is a unitary matrix.
The rotation operator $\hat{R}$ transforms the canonical basis as
\begin{eqnarray}
    U ~ \overset{\hat{R}}{\longrightarrow} ~ W = UD^{\intercal},  \label{unitaryW}
\end{eqnarray}
where $D^{\intercal}$ is the transpose of the Wigner-$D$ matrix for single-particle states.
Using Eqs. (\ref{cooper}), (\ref{bp}), and (\ref{unitaryW}), the norm between two different broken-pair states can be written as
\begin{widetext}
\begin{eqnarray}
		\langle \alpha | \hat{R} |  \beta^{\prime} \rangle   &=&  \sum_{\substack{i_1 \cdots i_N  \\ j_1 \cdots j_N  }}  
		v_{i_1} \cdots v_{i_N} v^{\prime}_{j_1} \cdots v^{\prime}_{j_N} 
		\sum_{k_1 \cdots k_n} \delta_{ k_1 \cdots k_n}^{ j_1 \tilde{j}_1 \cdots  j_N \tilde{j}_N  \beta_1 \cdots \beta_{\nu}}   X_{i_1  k_1} X_{\tilde{i}_1  k_2} \cdots X_{i_N k_{2N-1}} X_{\tilde{i}_N k_{2N}}  X_{\alpha_1 k_{2N+1}} \cdots X_{\alpha_{\nu} k_{n}}  \nonumber\\
		&=&  (N!)^2 \sum_{\substack{\{i_1 \cdots i_N\}\\\{j_1 \cdots j_N\}}} v_{i_1} \cdots v_{i_N} v^{\prime}_{j_1} \cdots v^{\prime}_{j_N} \det\left( X_{\{i\alpha,j\beta\}} \right),\label{eq6}
\end{eqnarray}
\end{widetext}
\noindent where $X \equiv U^{*} W^{\prime \intercal} $, and $\delta_{ a_1 \cdots a_n}^{b_1 \cdots b_n} $ is the generalized Kronecker delta: $\delta_{ a_1 \cdots a_n}^{b_1 \cdots b_n} = +1$ ($-1$) if $a_1, \cdots, a_n$ are distinct and form an even (odd) permutation of $b_1, \cdots, b_n$, and $=0$ otherwise.
In obtaining the second line, the sum over $k_1,\ldots,k_n$ involving the generalized Kronecker delta and the product of matrix elements of $X$ yields a minor of $X$.
$\det\left( X_{\{i\alpha,j\beta\}} \right)$ denotes the minor, i.e., the determinant of the submatrix formed by selecting rows $ i_1, \tilde{i}_1, \cdots,  i_N, \tilde{i}_N,  \alpha_1, \cdots,\alpha_{\nu}$ and columns $ j_1, \tilde{j}_1, \cdots,  j_N, \tilde{j}_N,  \beta_1, \cdots, \beta_{\nu}$ of $X$.
This minor represents the overlap between the Slater determinants formed from the selected canonical orbitals of the states $\ket{\alpha}$ and $\hat{R}\ket{\beta}^{\prime}$.
The summation $\sum_{\{i_1 \cdots i_N\}}$ runs over all $N$-element combinations from $ \{1, \cdots,\Omega \} $, 
reflecting the pair correlation.
Eq. (\ref{eq6}) and related formulas can also be derived using Grassmann algebra \cite{Grassmann1,Grassmann2}.
The details of this alternative derivation are omitted here for brevity.

The norm $\langle \alpha | \hat{R} |  \beta^{\prime} \rangle$ can be expressed in an alternative form.
To derive this, I rewrite the Cooper pair creation operator in Eq.~(\ref{cooper}) as
\begin{eqnarray}
	\hat{P}^{^{\dagger}} \equiv \frac{1}{2} \sum_{i j} \mathcal{P}_{ij} \hat{a}^{\dagger}_{i} \hat{a}^{\dagger}_{j}, \label{general}
\end{eqnarray}
where the matrix $\mathcal{P} = U^{\intercal} \mathcal{V} U$, and
\begin{eqnarray}
	\mathcal{V} = \left( 
	\begin{array}{ccccccc}
		0 & v_1 & &&\multicolumn{3}{c}{\multirow{4}*{ \Large0 } } \\
		-v_1 & 0 & && \\
		& & 0 & v_2 &   \\
		&  & -v_2 & 0 &  \\
		\multicolumn{4}{c}{\multirow{3}*{ \Large0 } }& \ddots& &  \\
		&&&& & 0& v_{\Omega} \\
		&&&&& -v_{\Omega}   & 0
	\end{array}
	\right).
\end{eqnarray}
The equivalence of Eqs. (\ref{cooper}) and (\ref{general}) can be obtained from Refs.~\cite{Youla1961,10.1063/1.1724294}.
Under rotation, the Cooper pair creation operator becomes
\begin{eqnarray}
  \hat{P}^{^{\dagger}}  \overset{\hat{R}}{\longrightarrow}  \frac{1}{2} \sum_{i j} \mathcal{Q}_{ij} \hat{a}^{\dagger}_{i} \hat{a}^{\dagger}_{j},     \label{Q}
\end{eqnarray}
where $\mathcal{Q} = W^{\intercal} \mathcal{V} W$. 
The norm is then expressed as
\begin{widetext}
\begin{eqnarray}
  	\langle \alpha | \hat{R} |  \beta^{\prime} \rangle &=& \frac{1}{2^{2N}} \sum_{\substack{i_1 \cdots i_{2N}   \\ j_1 \cdots j_{2N}  } }  
			\mathcal{P}^*_{i_1 i_2}  \cdots \mathcal{P}^*_{i_{2N-1} i_{2N}} 
		\mathcal{Q}^{\prime}_{j_1 j_2}  \cdots \mathcal{Q}^{\prime}_{j_{2N-1} j_{2N}}    	\sum_{\substack{  a_1 \cdots a_{\nu} \\  b_1 \cdots b_{\nu}} } 
		\delta_{i_1 \cdots i_{2N} a_1 \cdots a_{\nu}}^{ j_1 \cdots j_{2N} b_1 \cdots b_{\nu}}    
	    U^*_{\alpha_1 a_1} \cdots U^*_{\alpha_{\nu} a_{\nu}}  W^{\prime}_{\beta_1 b_1} \cdots W^{\prime}_{\beta_{\nu} b_{\nu}} 
		\nonumber\\
	& =& (N!)^2 \sum_{ \mathbb{C}_n} \sum_{ \substack{\{a_1 \cdots a_{\nu}\} \subseteq \mathbb{C}_n \\ \{b_1 \cdots b_{\nu}\}  \subseteq \mathbb{C}_n }   }  
	    {\rm sgn}(ia) ~ {\rm sgn}(jb)    	\det\left( U^*_{\{\alpha,a\}} \right) \det\left( W^{\prime}_{\{\beta,b\}} \right) {\rm pf} \left(\mathcal{P}^*_{\{i\}} \right)  {\rm pf}\left( \mathcal{Q}^{\prime}_{\{j\}} \right)
		 , \label{eq5}
\end{eqnarray}
\end{widetext}
\noindent where $ \mathbb{C}_n$ represents an $n$-element combination from $\{1, \cdots,\omega \}$, and the summation $\sum_{ \mathbb{C}_n}$ runs over all such combinations.
Each combination $ \mathbb{C}_n$ must satisfy
\begin{eqnarray}
\mathbb{C}_n &=& \{i_1 ,\cdots, i_{2N} ,a_1 ,\cdots, a_{\nu} \} \nonumber\\
         &=&   \{ j_1, \cdots, j_{2N}, b_1, \cdots ,b_{\nu}  \} \subseteq \{1, \cdots,\omega \}.
\end{eqnarray}
The term ${\rm sgn}(ia)$ denotes the sign of the permutation $(i_1 \cdots  i_{2N} a_1  \cdots  a_{\nu})$,
and ${\rm pf}\left( \mathcal{Q}^{\prime}_{\{j\}} \right)$ denotes the Pfaffian of the submatrix of $\mathcal{Q}^{\prime}$ whose rows and columns are indexed by $j_1, \cdots, j_{2N}$.
For $\nu=0$, Eq.~(\ref{eq5}) reduces to the Pfaffian expression for the overlap between pair condensates~\cite{MIZUSAKI2018237}.
It is worth mentioning that in the pioneering work of Ref. \cite{PhysRevC.79.021302}, Robledo introduced a Pfaffian formula to resolve the sign problem in HFB wave functions. 
Following this approach, substantial efforts have been devoted to improving HFB calculations using Pfaffian formulas, as demonstrated in Refs.~\cite{PhysRevC.84.014307,PhysRevLett.108.042505,GAO2014360,MIZUSAKI2018237,PhysRevLett.126.172501}.

Next, I present a formula for the matrix element of a many-body operator between broken-pair states.
An $m$-body operator is defined as
\begin{eqnarray}
	\hat{O}_m \equiv \hat{a}^{\dagger}_{f_1} \hat{a}^{\dagger}_{f_2} \cdots \hat{a}^{\dagger}_{f_m} \hat{a}_{g_m} \hat{a}_{g_{m-1}} \cdots \hat{a}_{g_1}. \label{operator}
\end{eqnarray}
Using Eqs. (\ref{cooper}), (\ref{bp}), (\ref{unitaryW}), and (\ref{operator}), the matrix element of the $m$-body operator is expressed in a compact form:
\begin{widetext}
\begin{eqnarray}
	   \langle \alpha | \hat{O}_m \hat{R} |  \beta^{\prime} \rangle &=& \langle \alpha | \hat{a}^{\dagger}_{f_1} \hat{a}^{\dagger}_{f_2} \cdots \hat{a}^{\dagger}_{f_m} \hat{a}_{g_m} \hat{a}_{g_{m-1}} \cdots \hat{a}_{g_1} \hat{R} |  \beta^{\prime} \rangle  \nonumber\\
		&=&  \frac{1}{(n-m)!} \sum_{\substack{i_1 \cdots i_N  \\ j_1 \cdots j_N  }}  
		v_{i_1} \cdots v_{i_N} v^{\prime}_{j_1} \cdots v^{\prime}_{j_N} 
		\sum_{\substack{a_1 \cdots a_m  c_1 \cdots c_{n-m}  \\ b_1 \cdots b_m d_1 \cdots d_{n-m}  }}  
		\delta_{ a_1 \cdots a_m c_1 \cdots c_{n-m} }^{ i_1 \tilde{i}_1 \cdots  i_N \tilde{i}_N  \alpha_1 \cdots \alpha_{\nu}}    
		~ \delta_{ b_1 \cdots b_m d_1 \cdots d_{n-m} }^{ j_1 \tilde{j}_1 \cdots  j_N \tilde{j}_N  \beta_1 \cdots \beta_{\nu}}    
		\nonumber\\
		&& \quad \times   U^*_{  a_1 f_1} \cdots U^*_{a_m f_m }  W^{\prime}_{ b_1 g_1} \cdots W^{\prime}_{ b_m g_m}  X_{c_1  d_1}   \cdots   X_{c_{n-m} d_{n-m}} 
		\nonumber\\
		&=&   (N!)^2 \sum_{\substack{\{i_1 \cdots i_N\}\\\{j_1 \cdots j_N\}}} 
		 v_{i_1} \cdots v_{i_N} v^{\prime}_{j_1} \cdots v^{\prime}_{j_N} 
		 \sum_{\substack{ \{a_1 \cdots a_m \} \\ \{b_1 \cdots b_m\}  }} 
	      {\rm sgn}(ac) ~ {\rm sgn}(bd) \det\left(U^*_{\{a,f\}}\right) \det\left(W^{\prime}_{\{b,g\}}\right) \det\left(X_{\{c,d\}}\right).\label{eq8}
\end{eqnarray}
\end{widetext}
In obtaining the second line, I have used the identity:
\begin{eqnarray}
	 \hat{ \mathbb{1}} = \sum_k \frac{1}{k!}  \sum_{i_1 \cdots i_k} \hat{a}^{\dagger}_{i_1} \hat{a}^{\dagger}_{i_2} \cdots \hat{a}^{\dagger}_{i_k}  \ket{}  \bra{} \hat{a}_{i_k} \hat{a}_{i_{k-1}} \cdots \hat{a}_{i_1}.
\end{eqnarray}

Similar to the norm formulas, an alternative expression for the matrix element of the $m$-body operator is presented. Using Eqs. (\ref{bp}), (\ref{unitaryW}), (\ref{general}) and (\ref{Q}), the matrix element can be written as
\begin{widetext}
\begin{eqnarray}
		 \langle \alpha | \hat{O}_m \hat{R} |  \beta^{\prime} \rangle &=&  \frac{1}{2^{2N}(n-m)!} \sum_{\substack{i_1 \cdots i_{2N}   \\ j_1 \cdots j_{2N}  } }  
		\mathcal{P}^*_{i_1 i_2}  \cdots \mathcal{P}^*_{i_{2N-1} i_{2N}} 
		\mathcal{Q}^{\prime}_{j_1 j_2}  \cdots \mathcal{Q}^{\prime}_{j_{2N-1} j_{2N}}     \sum_{\substack{  a_1 \cdots a_{\nu} \\  b_1 \cdots b_{\nu}  }}  \sum_{  d_1 \cdots d_{n-m} }   
		\delta_{f_1 \cdots f_m d_1 \cdots d_{n-m}}^{ i_1 \cdots i_{2N} a_1 \cdots a_{\nu}}    
	~	\delta_{g_1 \cdots g_m d_1 \cdots d_{n-m}}^{ j_1 \cdots j_{2N} b_1 \cdots b_{\nu}}    \nonumber\\
		&& \quad  \times ~
		U^*_{\alpha_1 a_1} \cdots U^*_{\alpha_{\nu} a_{\nu}}  W^{\prime}_{\beta_1 b_1} \cdots W^{\prime}_{\beta_{\nu} b_{\nu}} 
		\nonumber\\
		&=&  (N!)^2 \sum_{ \mathbb{D}_{n-m}} \sum_{ \substack{\{a_1 \cdots a_{\nu}\} \subseteq \mathbb{F}_n \\ \{b_1 \cdots b_{\nu}\}  \subseteq \mathbb{G}_n }   }  
	   {\rm sgn}(ia) ~ {\rm sgn}(jb) ~  {\rm sgn}(fd) ~  {\rm sgn}(gd)	  \det\left(U^*_{\{\alpha,a\}}\right) \det\left(W^{\prime}_{\{\beta,b\}}\right) {\rm pf}\left(\mathcal{P}^*_{\{i\}}\right)  {\rm pf}\left(\mathcal{Q}^{\prime}_{\{j\}}\right),\label{eq7}
\end{eqnarray}
\end{widetext}
\noindent where the set $\mathbb{D}_{n-m}$ is defined as
\begin{eqnarray}
	\mathbb{D}_{n-m} &=& \{d_1, \cdots, d_{n-m} \} \nonumber \\
	&\subseteq&  \left[ \{ 1, \cdots,\omega \} \setminus \left( \{ f_1, \cdots, f_m \} \cup \{ g_1, \cdots, g_m \}  \right) \right], \nonumber \\
\end{eqnarray}
and the sets $\mathbb{F}_{n}$ and $\mathbb{G}_{n}$ are defined as
\begin{eqnarray}
	\mathbb{F}_{n} &=& \mathbb{D}_{n-m} \cup \{ f_1, \cdots, f_m \},  \nonumber \\
	\mathbb{G}_{n} &=& \mathbb{D}_{n-m} \cup \{ g_1, \cdots, g_m \}.
\end{eqnarray}
Eqs. (\ref{eq7}) and (\ref{eq5}) are equivalent to Eqs. (\ref{eq8}) and (\ref{eq6}), respectively.

As a consistency check, I consider the $N$-pair condensate $\ket{\mathrm{c}}$ defined in Eq.~(\ref{pc}), in a doubly degenerate Hartree-Fock basis with the pairing Hamiltonian:
\begin{eqnarray}
	\hat{H}_{\rm p} \equiv \sum_{k} \varepsilon_k (\hat{a}^{\dagger}_{k} \hat{a}_{k} +\hat{a}^{\dagger}_{\tilde{k}} \hat{a}_{\tilde{k}} ) - \sum_{kl} g_{kl} \hat{a}^{\dagger}_{k} \hat{a}^{\dagger}_{\tilde{k}} \hat{a}_{\tilde{l}} \hat{a}_{l}.
\end{eqnarray}
Using Eqs. (\ref{eq6}) and (\ref{eq8}), the norm and Hamiltonian matrix element are 
\begin{eqnarray}
	\langle {\rm c} |  {\rm c} \rangle &=& (N!)^2 \sum_{\{i_1 \cdots i_N\}} v^2_{i_1} \cdots v^2_{i_N} ,  \label{overlap1}  \\
	\langle {\rm c} | \hat{H}_{\rm p} |  {\rm c} \rangle &=& (N!)^2 \sum_{\{i_1 \cdots i_{N-1}\}} v^2_{i_1} \cdots v^2_{i_{N-1}} \nonumber\\
	&& \times\sum_{kl \notin \{i_1 \cdots i_{N-1}\} }  v_{k} v_{l}(2\varepsilon_k \delta_{kl} - g_{kl}) .\label{H1}
\end{eqnarray}
For degenerate single-particle states with constant pairing strength ($\varepsilon_k \equiv \varepsilon$ and $g_{kl} \equiv g$), the pair coefficients $v_k$ become uniform, and the norm in Eq. (\ref{overlap1}) reaches its maximum value for a normalized pair: $ (N!)^2 \binom{\Omega}{N} / \Omega^N $.
This value is smaller than the norm of an $N$-boson condensate $N!$ due to the Pauli principle, and approaches $N!$ in the limit $\Omega\rightarrow\infty$.
Using Eq. (\ref{H1}), the energy in this case is $E= 2\varepsilon N - gN(\Omega-N+1)$, which recovers the result obtained from other approaches, such as quasi-spin theory \cite{quasispin}.



\section{Application of PNBCS to $^{70}$Ge and $^{69}$Ge}

One of the applications of the present formalism is the study of collective states using the angular-momentum-projected number-conserved BCS (PNBCS) method~\cite{PNBCS,PNBCS2025Letter,PNBCS2025RareEarth}. 
In this work, I consider both the even-even nucleus $^{70}$Ge and the
odd-mass nucleus $^{69}$Ge in the
$1p_{1/2}1p_{3/2}0f_{5/2}0g_{9/2}$ space above the $^{56}$Ni core, using the
JUN45 effective interaction \cite{jun45}. 
The PNBCS calculation is performed in three steps.
First, I solve the self-consistent Hartree--Fock (HF) equation with the Kramers time-reversal symmetry for $^{70}$Ge under the JUN45 interaction, without enforcing any other constraints like shapes or orientations. The solutions are called ``intrinsic'' HF states. One usually obtains multiple distinct HF solutions. 
Then, for a given HF basis I consider the pair condensate in Eq.~(\ref{pc}), where the Cooper pair coefficients are determined by solving the number-conserved BCS (NBCS) equation \cite{PNBCS,PhysRevC.96.034313,PhysRevC.99.014302}.
Finally, the energies and wave functions are obtained by angular momentum and parity projection, i.e., solving the Hill--Wheeler equation.

For $^{70}$Ge, the projected space mixes two pair-condensate structures obtained from an axially symmetric oblate
HF solution with $(\beta,\gamma)=(0.13,60^\circ)$ and a triaxial solution
with $(\beta,\gamma)=(0.25,36^\circ)$. The deformation parameters are
calculated following the convention of Ref.~\cite{PNBCS2025RareEarth}.
The Hill--Wheeler equation can be written as
\begin{equation}
\sum_{bK}{\mathcal H}^{J\pi}_{aK',bK}g^r_{bK}
=\epsilon_{J_r^\pi}\sum_{bK}{\mathcal N}^{J\pi}_{aK',bK}g^r_{bK},
\label{eqn:HillWheeler}
\end{equation}
where $a$ and $b$ label the two intrinsic HF configurations, and $\pi$ is the parity.
For comparison, full-configuration-interaction (FCI) shell-model calculations are
performed using BIGSTICK \cite{bigstick,johnson2018bigstick}, with the same
interaction and single-particle space.
To calculate electric-quadrupole transition strengths $B(E2)$, I use the effective charges $e_\pi=1.5e$ and $e_\nu=0.8e$. 

\begin{figure}[tbp]
\centering\includegraphics[width=\columnwidth]{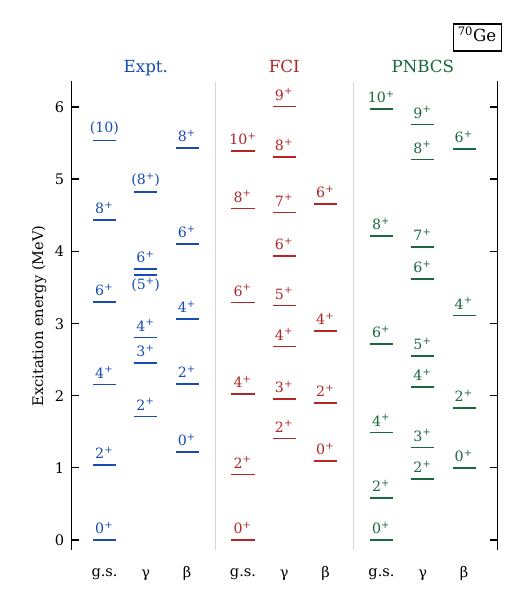}
\caption{Excitation energies of the ground, $\gamma$, and $\beta$ bands in $^{70}$Ge from experiment
\cite{GURDAL20161}, the FCI, and PNBCS. The experimental ground band shows the $8^+$ state with the larger $B(E2)$ to $6^+$ and its $10^+$ candidate; the experimental spin assignment $J=10$ remains uncertain. 
The experimental $\beta$ band is adopted from Ref.~\cite{GURDAL20161}. But the collective interpretation of the $4^+$, $6^+$, and $8^+$ states is uncertain, as discussed in the text.}
\label{fig1}
\end{figure}

\begin{table}[tbp]
\small\setlength{\tabcolsep}{2pt}
\caption{$B(E2)$ values in Weisskopf units (W.u.) in $^{70}$Ge.
$\Delta I=2$ transitions are listed for the ground and $\beta$ bands, and both $\Delta I=1$ and $\Delta I=2$ for the $\gamma$ band.
The experimental data are taken from Ref.~\cite{GURDAL20161} and the recent measurement of Ref.~\cite{Kowalewski2025Ge70}.
The calculations use effective charges $e_\pi=1.5e$ and $e_\nu=0.8e$. }
\label{tab70BE2}
\begin{ruledtabular}
\begin{tabular}{>{\raggedright\arraybackslash}p{35pt}lrrrr}
Band & $J_i\rightarrow J_f$ & Ref.~\cite{GURDAL20161} & Ref.~\cite{Kowalewski2025Ge70} & FCI & PNBCS \\
\hline
ground & $2^{+}\rightarrow0^{+}$ & $20.8^{+0.4}_{-0.4}$ & $20.8^{+0.4}_{-0.4}$ & 20.65 & 25.93 \\
 & $4^{+}\rightarrow2^{+}$ & $25^{+5}_{-5}$ & $30.0^{+4.4}_{-0.9}$ & 33.82 & 36.79 \\
 & $6^{+}\rightarrow4^{+}$ & $34^{+7}_{-7}$ & $25^{+4}_{-5}$ & 39.60 & 46.33 \\
 & $8^{+}\rightarrow6^{+}$ & $44^{+22}_{-22}$ &  & 26.03 & 47.70 \\
 & $10^{+}\rightarrow8^{+}$ &  &  & 23.08 & 45.95 \\
$\gamma$ & $3^{+}\rightarrow2^{+}$ & $<0.5^{a}$ &  & 41.57 & 48.31 \\
 & $4^{+}\rightarrow2^{+}$ & $31^{+11}_{-11}$ & $12^{+2}_{-1}$ & 10.15 & 13.53 \\
 & $4^{+}\rightarrow3^{+}$ &  &  & 5.89 & 3.38 \\
 & $5^{+}\rightarrow3^{+}$ &  &  & 21.02 & 25.22 \\
 & $5^{+}\rightarrow4^{+}$ &  &  & 23.70 & 25.37 \\
 & $6^{+}\rightarrow4^{+}$ & $27^{+9}_{-9}$ &  & 13.59 & 19.93 \\
 & $6^{+}\rightarrow5^{+}$ &  &  & 0.03 & 0.34 \\
 & $7^{+}\rightarrow5^{+}$ &  &  & 24.51 & 31.38 \\
 & $7^{+}\rightarrow6^{+}$ &  &  & 13.02 & 19.40 \\
 & $8^{+}\rightarrow6^{+}$ &  &  & 17.07 & 19.67 \\
 & $8^{+}\rightarrow7^{+}$ &  &  & 0.59 & 0.92 \\
 & $9^{+}\rightarrow7^{+}$ &  &  & 10.12 & 32.78 \\
 & $9^{+}\rightarrow8^{+}$ &  &  & 3.50 & 19.91 \\
$\beta$ & $2^{+}\rightarrow0^{+}$ &  & $14.21^{+0.65}_{-0.73}$ & 11.38 & 11.43 \\
 & $4^{+}\rightarrow2^{+}$ & $1.00^{+0.23}_{-0.23}$ &  & 10.59 & 13.81 \\
 & $6^{+}\rightarrow4^{+}$ &  &  & 4.22 & 13.96 \\
 & $8^{+}\rightarrow6^{+}$ &  &  &  &  \\
$\gamma$ to ground & $2^{+}\rightarrow0^{+}$ & $0.50^{+0.08}_{-0.08}$ & $0.32^{+0.01}_{-0.01}$ & 0.66 & 0.49 \\
 & $2^{+}\rightarrow2^{+}$ & $64^{+11}_{-11}$ & $20.2^{+1.3}_{-0.9}$ & 31.72 & 37.63 \\
 & $3^{+}\rightarrow2^{+}$ & $0.79^{+0.48}_{-0.17}$ &  & 1.35 & 1.02 \\
$\gamma$ to $\beta$ & $2^{+}\rightarrow0^{+}$ & $16^{+3}_{-3}$ & $9.9^{+0.7}_{-0.5}$ & 1.67 & 0.63 \\
$\beta$ to ground & $0^{+}\rightarrow2^{+}$ & $48^{+7}_{-7}$ & $31^{+1}_{-3}$ & 22.84 & 11.56 \\
 & $4^{+}\rightarrow2^{+}$ & $0.30^{+0.07}_{-0.07}$ &  & 0.05 & 0.27 \\
ground to $\gamma$ & $4^{+}\rightarrow2^{+}$ & $17^{+10}_{-10}$ & $12^{+1}_{-2}$ & 0.63 & 0.81 \\
$\beta$ to $\gamma$ & $4^{+}\rightarrow2^{+}$ & $0.063^{+0.016}_{-0.016}$ &  & 0.70 & 0.01 \\
\end{tabular}
\end{ruledtabular}
\parbox{\columnwidth}{\footnotesize $^{a}$The adopted value depends on
conflicting experimental $E2/M1$ mixing ratios; see the discussion in the text.}

\end{table}

Figure~\ref{fig1} compares the PNBCS and FCI spectra of $^{70}$Ge with experiment \cite{GURDAL20161}, and Table~\ref{tab70BE2} lists the corresponding $B(E2)$ values.
Here only the ground, $\gamma$, and $\beta$ bands are presented.
The collective bands are identified using relatively larger $B(E2)$ values. 
I exemplify this with the ground-band $8^+$ state. 
Ref.~\cite{GURDAL20161} reports that the $8^+$ state at $4.431$~MeV feeds the ground-band $6^+$ state with $B(E2)=44\pm22$~W.u., while the yrast $8^+$ state at $4.204$~MeV feeds the $6^+$ with only $B(E2)=6.7\pm1.7$~W.u. 
Thus the former is assigned as the ground-band $8^+$ state in this work.
Similarly, in FCI the $8^+$ state at $4.597$~MeV feeds the ground-band $6^+$ with $B(E2)=26$~W.u., while the yrast $8^+$ at $3.927$~MeV feeds it with only $B(E2)=6.3$~W.u.
The PNBCS calculation shows that the ground and $\gamma$ bands arise predominantly from the triaxial HF configuration, whereas the $\beta$ band is mainly associated with the oblate HF configuration.
It is interesting to see that while the $\beta$-band $0^+$ state is $\sim 1$~MeV higher than the ground-band $0^+$, the oblate HF solution lies $0.08$~MeV lower than the triaxial one. 
The ground-band $0^+$ state gains more energy from angular-momentum projection and configuration mixing. 


The experimental $\gamma$ band shows odd--even staggering with close $3^+$--$4^+$ and $5^+$--$6^+$ pairs, indicating $\gamma$-soft character.
The FCI and PNBCS spectra instead show the opposite staggering regularity with $2^+$--$3^+$, $4^+$--$5^+$, and $6^+$--$7^+$ pairs; the staggering is much weaker in FCI.
Nevertheless, the FCI and PNBCS calculations show similar patterns of in-band $B(E2)$ values, with stronger $\Delta I=1$ $E2$ transitions
from odd-spin states and suppressed ones from even-spin states (see Table~\ref{tab70BE2}).
For the $3^+_\gamma\rightarrow2^+_\gamma$ transition, Ref.~\cite{GURDAL20161} adopts the mixing ratio $\delta=+0.04\pm0.08$ measured by Robinson \textit{et al.}~\cite{Robinson1977Ge70}, corresponding to predominantly $M1$ transition.
The later measurement of Sugawara \textit{et al.} \cite{Sugawara2010Ge70} also regarded it as a dipole transition.
However, the experimental result of Dostemesova \textit{et al.}~\cite{Dostemesova1987Ge}, listed as an alternative in Ref.~\cite{GURDAL20161}, gives $\delta=+3.5\pm0.9$ and implies predominantly $E2$ transition.
Thus the adopted $B(E2)<0.5$~W.u. remains uncertain.
Using the calculated $B(E2)$ and $B(M1)$ values together with the experimental $\gamma$ energy of $0.744$~MeV, I obtain $|\delta|=0.97$ and $1.06$ for FCI and PNBCS, respectively. 
The $B(M1)$ calculations use the free-nucleon orbital $g$ factors and spin $g$ factors with a quenching factor of $0.7$.
Both calculations predict comparable $M1$ and $E2$ contributions.

The experimental $\beta$ band in Fig.~\ref{fig1} is adopted from Ref.~\cite{GURDAL20161}. The $4^+$ state at $3.059$~MeV decays through the $2^+$ at $2.157$~MeV to the excited $0^+$, which was confirmed in Ref.~\cite{HaringKaye2018}. 
However, the measured $B(E2;4^+\rightarrow2^+)=1.00\pm0.23$~W.u. \cite{GURDAL20161} is much smaller than the FCI and PNBCS values of $10.59$ and $13.81$~W.u., respectively. This weak transition does not support a simple collective interpretation for the $4^+$.
The extension of this band to the $6^+$ and $8^+$ at $4.104$ and $5.436$~MeV was not confirmed in Ref.~\cite{HaringKaye2018}.
Among the 31 FCI $6^+$ states below $6$~MeV, the state at $4.661$~MeV has the largest $B(E2;6^+\rightarrow4^+)=4.22$~W.u. to the $\beta$-band $4^+$ state at $2.898$~MeV.
The PNBCS $6^+$ state at $5.422$~MeV feeds its $\beta$-band $4^+$ state with $13.96$~W.u. 
Both calculated $6^+$ states lie above the proposed experimental $6^+$.
Further experimental study is required.

\begin{figure*}[tbp]
\centering\includegraphics[width=\textwidth]{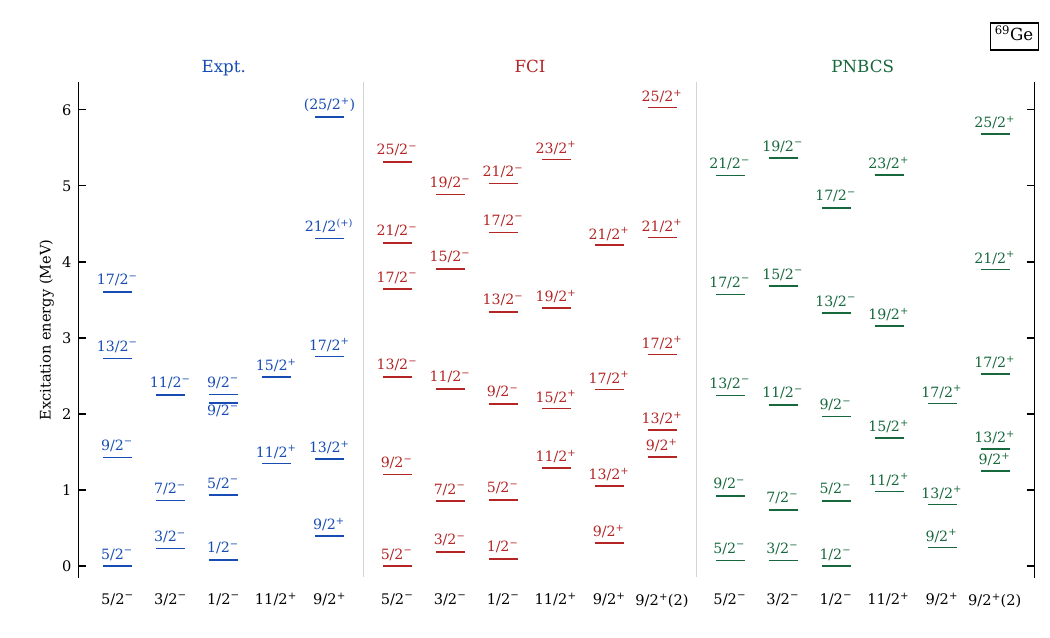}
\caption{Six selected collective bands in $^{69}$Ge:
the $5/2^-$, $3/2^-$, $1/2^-$, $11/2^+$, first $9/2^+$, and second $9/2^+$ bands. 
No experimental level is assigned to the second $9/2^+$ band.
The two experimental $9/2^-$ states in the $1/2^-$ band are alternative candidates.
Experimental data are from Ref.~\cite{NESARAJA2026}. }
\label{fig69main}
\end{figure*}

\begin{table}[tbp]
\small\setlength{\tabcolsep}{2pt}
\caption{$B(E2)$ (W.u.) in $^{69}$Ge. 
The experimental data are taken from Ref.~\cite{NESARAJA2026}.}
\label{tab69BE2}
\begin{ruledtabular}
\begin{tabular}{llrrr}
Band & $J_i\rightarrow J_f$ & Expt. & FCI & PNBCS \\
\hline
$5/2^{-}$ & $9/2^{-}\rightarrow5/2^{-}$ & $8.5^{+2.4}_{-1.6}$ & 14.31 & 12.64 \\
 & $13/2^{-}\rightarrow9/2^{-}$ &  & 18.76 & 13.83 \\
 & $17/2^{-}\rightarrow13/2^{-}$ & $19^{+9}_{-5}$ & 18.77 & 39.30 \\
 & $21/2^{-}\rightarrow17/2^{-}$ &  & 15.14 & 41.07 \\
 & $25/2^{-}\rightarrow21/2^{-}$ &  & 21.59 &  \\
$3/2^{-}$ & $7/2^{-}\rightarrow3/2^{-}$ & $14^{+5}_{-5}$ & 6.48 & 11.24 \\
 & $11/2^{-}\rightarrow7/2^{-}$ & $13^{+10}_{-5}$ & 13.00 & 20.27 \\
 & $15/2^{-}\rightarrow11/2^{-}$ &  & 10.29 & 22.73 \\
 & $19/2^{-}\rightarrow15/2^{-}$ &  & 6.32 & 30.28 \\
$1/2^{-}$ & $5/2^{-}\rightarrow1/2^{-}$ & $13^{+10}_{-4}$ & 10.20 & 12.14 \\
 & $9/2^{-}\rightarrow5/2^{-}$ &  & 12.11 & 12.41 \\
 & $13/2^{-}\rightarrow9/2^{-}$ &  & 18.50 & 14.53 \\
 & $17/2^{-}\rightarrow13/2^{-}$ &  & 8.86 & 10.14 \\
 & $21/2^{-}\rightarrow17/2^{-}$ &  & 12.22 &  \\
$11/2^{+}$ & $15/2^{+}\rightarrow11/2^{+}$ & $10.7^{+4.8}_{-2.6}$ & 15.39 & 17.47 \\
 & $19/2^{+}\rightarrow15/2^{+}$ &  & 15.17 & 20.49 \\
 & $23/2^{+}\rightarrow19/2^{+}$ &  & 11.61 & 19.30 \\
$9/2^{+}$ & $13/2^{+}\rightarrow9/2^{+}$ & $20.5^{+1.3}_{-1.3}$ & 17.52 & 16.53 \\
 & $17/2^{+}\rightarrow13/2^{+}$ & $12.8^{+2.6}_{-1.9}$ & 15.23 & 18.12 \\
 & $21/2^{+}\rightarrow17/2^{+}$ &  & 11.87 &  \\
 & $25/2^{+}\rightarrow21/2^{+}$ &  &  &  \\
$9/2^{+}$ (2) & $13/2^{+}\rightarrow9/2^{+}$ &  & 18.39 & 23.00 \\
 & $17/2^{+}\rightarrow13/2^{+}$ &  & 23.38 & 27.68 \\
 & $21/2^{+}\rightarrow17/2^{+}$ &  & 27.85 & 23.44 \\
 & $25/2^{+}\rightarrow21/2^{+}$ &  & 18.72 & 35.79 \\
\end{tabular}
\end{ruledtabular}
\end{table}

For $^{69}$Ge, the many-body basis is constructed by adding one neutron to the two pair condensates of $^{68}$Ge.
These pair condensates are obtained by NBCS optimization in the bases of two triaxial HF solutions with $(\beta,\gamma)=(0.18,38^\circ)$ and
$(0.27,18^\circ)$, respectively, which describe the shape coexistence phenomenon in $^{68}$Ge well~\cite{PNBCS}.
Thus the many-body basis for $^{69}$Ge corresponds to $\nu=1$ in Eq.~(\ref{bp}).
This nucleus tests the capability of PNBCS to describe low-lying collective states in odd-mass nuclei.

Figure~\ref{fig69main} compares the calculated collective bands in $^{69}$Ge with experiment \cite{NESARAJA2026}. 
Both FCI and PNBCS identify six low-lying bands: the $5/2^-$, $3/2^-$, $1/2^-$, $11/2^+$, first $9/2^+$, and second $9/2^+$ bands.
No experimental levels are assigned to the second $9/2^+$ band because the transition data are insufficient.
For the $5/2^-$ band, both calculations predict the $17/2^-$ energy close to the experimental value of $3.605$~MeV.
The measured $B(E2; 11/2^-\rightarrow7/2^-)$ supports the assignment of the $11/2^-$ state to the $3/2^-$ band (see Table~\ref{tab69BE2}).
In the experimental $1/2^-$ band, the $9/2^-$ states at $2.149$ and $2.258$~MeV decay to the $5/2^-$ state by $E2$ transitions. Their $B(E2)$ values are unknown, so they are shown as alternative $9/2^-$ candidates. 
For the experimental $11/2^+$ band, the measured $B(E2; 15/2^+\rightarrow11/2^+)$ supports the assignment of the $15/2^+$ state.
The available data do not establish higher-spin candidates of these experimental bands.


The experimental, FCI, and PNBCS results agree well for the lower-lying excitation energies and $B(E2)$ values (see Table~\ref{tab69BE2}).
Discrepancies appear at higher levels.
In FCI, the $21/2^+\rightarrow17/2^+$ $B(E2)$ values support separate extensions of the two $9/2^+$ bands.
In PNBCS, the $21/2^+$ state at $3.898$~MeV feeds the $17/2^+$ states of the first and second $9/2^+$ bands with $B(E2)=12.29$ and $23.44$~W.u., respectively.
The larger value supports its assignment to the second band.
For the $3/2^-$ band, the FCI $19/2^-$ states at $4.882$ and $4.859$~MeV decay to the $15/2^-$ state with $B(E2)=6.32$ and $4.38$~W.u., respectively.
The state with the larger $B(E2)$ is shown as a tentative band candidate in Fig.~\ref{fig69main}.
PNBCS gives larger in-band $B(E2)$ values than FCI for some higher-spin states.
Configuration mixing beyond the adopted configuration space might redistribute the $E2$ strength among nearby states.
The model space can be enlarged by including more intrinsic pair structures and/or higher-seniority configurations.

To examine the structures of the low-lying collective bands in $^{69}$Ge, I label the two intrinsic pair condensates of $^{68}$Ge as even-even cores 1 and 2, in the order given above.
For each core $i$, I couple the unpaired neutron to that pair condensate and project onto the chosen $J^\pi$.
The subspace spanned by the projected states is denoted by $\mathcal{V}_{J^\pi,i}$.
For a normalized PNBCS eigenstate of $^{69}$Ge, $\ket{\Psi}$,
I calculate
\begin{eqnarray}
	W_{i}=\bra{\Psi}\hat P_{J^\pi,i}\ket{\Psi},
\end{eqnarray}
where $\hat P_{J^\pi,i}$ is the projection operator onto the subspace $\mathcal{V}_{J^\pi,i}$.
Thus $W_i$ estimates how well $\ket{\Psi}$ can be represented by configurations constructed from intrinsic core $i$.
Note that the two core subspaces $\mathcal{V}_{J^\pi,1}$ and $\mathcal{V}_{J^\pi,2}$ are generally nonorthogonal, so $W_1+W_2$ need not equal unity.

The calculation shows a change of core overlaps with increasing spin in the $5/2^-$ band.
The $5/2^-$ and $9/2^-$ states have larger overlaps with core 1,
with $(W_1,W_2)=(0.953,0.529)$ and $(0.831,0.590)$, respectively.
$(W_1,W_2)$ becomes $(0.471,0.981)$ at $13/2^-$ and
$(0.424,0.993)$ at $17/2^-$.
The similar feature appears in the $3/2^-$ and $1/2^-$ bands.
For the $3/2^-$ band, $(W_1,W_2)$ changes from $(0.898,0.546)$ at the bandhead to $(0.416,0.967)$ at $15/2^-$.
For the $1/2^-$ band, it changes from $(0.959,0.477)$ at the bandhead to $(0.408,0.955)$ at $17/2^-$.
The lower-spin positive-parity states have comparable overlaps with both core spaces. 
For example, for the bandheads of the $11/2^+$, first $9/2^+$, and second $9/2^+$ bands, $(W_1,W_2)=(0.726,0.811)$, $(0.759,0.889)$, and $(0.610,0.796)$, respectively.
These states are not fully described using either core alone, indicating that mixing of the two core configurations is important.

\section{\texorpdfstring{Summary}{Summary}}

To summarize, this work presents closed-form formulas for the norm and matrix elements of many-body operators between number-conserved pairing states.
The bra and ket can contain unpaired particles and have different canonical bases.
The expressions are derived using the generalized Kronecker delta and written in terms of minors and Pfaffians.
The formalism is applied to the study of low-lying collective bands in $^{70}$Ge and $^{69}$Ge using PNBCS with the JUN45 effective interaction, demonstrating its applicability to systems with and without unpaired particles.
For $^{70}$Ge, the calculated ground and $\gamma$ bands arise from the triaxial intrinsic pair condensate, while the $\beta$ band corresponds to the axially symmetric oblate structure.
For $^{69}$Ge, the calculated low-spin negative-parity states can be described mainly by coupling the unpaired neutron to the intrinsic even-even core with $(\beta,\gamma)=(0.18,38^\circ)$, while mixing with configurations based on the other core with $(\beta,\gamma)=(0.27,18^\circ)$ is important for the positive-parity bandheads.

\begin{acknowledgments}
The author thanks Prof. Y. Lu and Y. X. Yu for checking some of the test
calculations, and Prof. Calvin W. Johnson, Prof. Y. M. Zhao,
and Prof. C. Qi for valuable discussions.
This research was supported by the National Natural Science
Foundation of China under Grant Nos.~12322506, 12535009,
and 12075169, and by the Fundamental Research Funds for the Central Universities.
\end{acknowledgments}

\bibliographystyle{apsrev4-1}
\bibliography{gjfu-bib}

\end{document}